\documentclass[traditabstract]{aa}

\usepackage{graphicx}
\usepackage{txfonts}
\usepackage{natbib}
\usepackage{xspace}
\usepackage{color}
\usepackage{hyperref}
\bibpunct{(}{)}{;}{a}{}{,}

\newcommand{\pycs}{{\tt PyCS}\xspace}
\newcommand{\edt}{{\widetilde{\Delta t_i}\xspace}}

\newcommand{\spl}{{\tt spl}\xspace}
\newcommand{\sdi}{{\tt sdi}\xspace}

\renewcommand{\sec}{{\tt sec}\xspace}
\newcommand{\secpla}{{\tt secpla}\xspace}

\newcommand{\splsec}{{\tt spl-sec}\xspace}
\newcommand{\sdisec}{{\tt sdi-sec}\xspace}

\newcommand{\stageone}{Stage I\xspace}
\newcommand{\stagetwo}{Stage II\xspace}

\begin{document}

\title{COSMOGRAIL: the COSmological MOnitoring of \\GRAvItational Lenses}
\subtitle{XV. Assessing the achievability and precision of time-delay measurements}
\titlerunning{COSMOGRAIL XV -- Time Delay Challenge}

\author{
V. Bonvin\inst{\ref{epfl}} \and
M. Tewes\inst{\ref{bonn}}  \and
F. Courbin\inst{\ref{epfl}} \and
T. Kuntzer\inst{\ref{epfl}} \and
D. Sluse\inst{\ref{iagul}} \and
G. Meylan\inst{\ref{epfl}}
}

\institute{
Laboratoire d'astrophysique, Ecole Polytechnique F\'ed\'erale de Lausanne (EPFL), Observatoire de Sauverny, 1290 Versoix, Switzerland \label{epfl}
\and
Argelander-Institut f\"ur Astronomie, Auf dem H\"ugel 71, D-53121 Bonn, Germany \label{bonn}
\and
Institut d’Astrophysique et de G\'eophysique, Universit\'e
de Li\`ege, All\'ee du 6 Ao\^ut 17, B5c, 4000 Li\`ege, Belgium \label{iagul}
}
\date{\today}

\abstract{COSMOGRAIL is a long-term photometric monitoring of gravitationally lensed quasars aimed at implementing Refsdal's time-delay method to measure cosmological parameters, in particular $H_0$. Given the long and well sampled light curves of strongly lensed quasars, time-delay measurements require numerical techniques whose quality must be assessed. To this end, and also in view of future monitoring programs or surveys such as the LSST, a blind signal processing competition named Time Delay Challenge 1 (TDC1) was held in 2014. The aim of the present paper, which is based on the simulated light curves from the TDC1, is double. First, we test the performance of the time-delay measurement techniques currently used in COSMOGRAIL. Second, we analyse the quantity and quality of the harvest of time delays obtained from the TDC1 simulations. To achieve these goals, we first discover time delays through a careful inspection of the light curves via a dedicated visual interface. Our measurement algorithms can then be applied to the data in an automated way. We show that our techniques have no significant biases, and yield adequate uncertainty estimates resulting in reduced $\chi^2$ values between 0.5 and 1.0. We provide estimates for the number and precision of time-delay measurements that can be expected from future time-delay monitoring campaigns as a function of the photometric signal-to-noise ratio and of the true time delay. We make our blind measurements on the TDC1 data publicly available.
}


\maketitle

\section{Introduction}

The methods used to constrain current cosmological models all benefit from independent measurements of the local value of the Hubble parameter, H$_0$ \citep[see e.g. Fig. 48 of][]{Weinberg2013}. One way of achieving a measurement of H$_0$ is based on time delays in strong gravitational lens systems. The method, first suggested by \citet{Refsdal1964}, proposes measuring the differences in the travel time of photons coming from multiple images of a distant source, such as a quasar. This time delay, $\Delta t$, is connected to the overall matter distribution responsible for the lensing effect, and to the time-delay distance $D_{\Delta t}$ to the lens, i.e. $H_0$, with some sensitivity to curvature and dark energy as well \citep[e.g.,][]{Suyu2014}.

Exploiting this relationship to constrain $H_0$ and cosmology in general requires both an accurate mass model for the lens and accurate time delay measurements \citep[see e.g.,][]{Suyu2012, Linder2011, Moustakas2009}. Modelling the lens mass in an unbiased way is difficult and prone to degeneracies known as the mass-sheet degeneracy \citep[e.g.,][]{SS2013} and, more generally, the source plane transformation described in \citet{SS2014}. The effect of lens model degeneracies can be mitigated by combining astrometric information from high resolution imaging, measurements of the lens dynamics, priors on the mass density profile of the lens, and an analysis of structures along the line of sight \citep[e.g.,][]{Suyu2014, Greene2013, Fadely2010, Treu2002, Falco1997, Keeton1997}. Integral field spectroscopy coupled with the adaptive optics that is becoming available on the VLT and at the Keck observatory will be essential in this respect. One of the key ingredient for the method to work at all is, however, the quality of the time-delay measurements, which is the focus of the present work.

In practice, measuring time delays is achievable if the lensed source is photometrically variable. Gravitationally lensed quasars are ideal targets: the quasars can show variability accessible by moderately sized ground-based optical telescopes on timescales of a few days, while the time delays resulting from galaxy-scale lenses are typically of the order of weeks to months \citep[see, e.g.,][]{Oguri2010}. The intrinsic light curve of the quasar is seen shifted by the relative time delays in the different lensed images. However, this simple situation is often contaminated: microlensing by stars in the lensing galaxy introduces extrinsic variability in the individual light curves with an amplitude sometimes comparable with that of the intrinsic variability of the quasar. To yield competitive cosmological constraints, reliable time-delay measurements with percent-level precision are needed \citep{Treu2010}. An efficient implementation of these measurements has long been hampered by how difficult it is to obtain photometric data for periods of many years at a desirable cadence, which must be close to 1 epoch per few days \citep{Eigenbrod2005}.

COSMOGRAIL is a monitoring program targeting more than 30 strongly lensed quasars using meter-class telescopes, with a cadence of 3 days for the most interesting systems. Recent results include light curves and time-delay measurements that are accurate to within a few percent points in HE~0435$-$1223 \citep{Courbin2011}, SDSS~J1001$+$5027 \citep{RathnaKumar:2013eu} and in RX~J1131$-$1231 \citep{Tewes:2013iz}. To measure these time delays, we developed and implemented several algorithms in the form of a COSMOGRAIL curve-shifting toolbox named \pycs, described in \citet{pycs}.

In the fall of 2013, a first blind time-delay measurement competition named Time Delay Challenge 1 (TDC1) was proposed to the community by \citet{Dobler2013}. The two main goals of this open challenge were (1) to homogeneously assess the quality of time-delay measurement algorithms on a common set of realistic synthetic light curves, and (2) to obtain some quantitative informations on the impact of observing strategy (cadence, season length, campaign length) on time-delay measurements. We took part in this challenge and submitted time-delay estimates using the team name \emph{PyCS}. \citet{Liao2014} give a summary of the results from all TDC1 participating teams, as well as some general conclusions. The present paper is complementary to \citet{Liao2014}. It focuses on the \pycs\ methods that we also apply to real light curves, and hence assesses the quality and reliability of the COSMOGRAIL time-delay measurements.

To evaluate our existing techniques with the thousands of light curves of TDC1 under conditions similar to the analysis of COSMOGRAIL data, we separated the problem of time-delay measurement of a light curve pair into two successive stages. 
\begin{description}
\item[\stageone :] we first attempt to \emph{discover} a plausible time delay, without trying to measuring it precisely. We also evaluated how confident we were that the proposed approximate solution is close to the true time delay and not a catastrophic failure. Owing to the limited length of the monitoring seasons, the randomness of quasar variability, noise and microlensing, this was not possible for every light curve pair of TDC1 or a real monitoring campaign. We note that in the case of TDC1 we had no prior information on the time delay to look for, as we had no knowledge of the mass distribution in the lensing galaxy. Only the light curves themselves were used. 

\item[\stagetwo :] for those systems for which \stageone was successful, we then focused on accurately estimating the time delay and associated uncertainty with the \pycs techniques, constraining the algorithms to a delay range around the solution from \stageone. As the \pycs \stagetwo methods did not rely on a physical model of the light curves, they would not be able to deal adequately with comparing odds among very different solutions.
\end{description}
This two-stage structure is of general interest beyond \pycs, as the stages concern discernible aspects of the time-delay measurement problem. \stageone deals with the quantity and the purity of time-delay measurements, while \stagetwo deals with their actual accuracy.

The paper is structured as follows. Section \ref{sec:tdc} summarizes the data from the Time Delay Challenge 1 and the metrics used to evaluate techniques. In Sect. \ref{sec:stage1}, we present the approaches we took to address \stageone, while Sect. \ref{sec:stage2} presents the \stagetwo algorithms. In Sect. \ref{sec:results} we discuss the results, expanding on the analysis of \citet{Liao2014}, and we conclude in Sect. \ref{sec:conclusion}.

\section{Time Delay Challenge 1}
\label{sec:tdc}

The mock data used in this work are synthetic quasar light curves made publicly available in the context of the the Time Delay Challenge 1 (TDC1) proposed by \citet{Dobler2013}. These data mimic the photometric variations seen in real gravitationally lensed quasars, with different time sampling, number of seasons, and season length. The curves are generated with simple yet plausible noise properties, and include microlensing variability. The dataset is split into five ``rungs'' or stages that simulate different observational strategies, each rung consisting of 1024 pairs of light curves. The rungs randomly mix microlensing, noise properties, and variability but differ in time sampling, number of seasons, and season length. These differences are listed in Table~1 of \citet[][hereafter TDC1 paper]{Liao2014}.

Participants to the TDC1 were asked to provide their best point estimate $\edt$ and associated 1-$\sigma$ uncertainty $\delta_i$ for as many pairs of curves as possible. The submitted entries to the challenge were then compared to the true input delays, and evaluated using simple metrics probing different properties of the estimates. The details of how the simulations were set up, as well as a discussion of these metrics are given in \citet{Dobler2013}. Results obtained by the blind submissions of the different participating teams are summarized in the TDC1 paper, including our submissions denoted ``\pycs''. For completeness, we briefly summarize the four main metrics:

\begin{enumerate}

\item The fraction $f$ of submitted time delay estimates,
\begin{equation}\label{eq:f}
f = N_{\rm submit}/N,
\end{equation}
where $N_{\rm submit}$ is the number of measured time delays and $N$ is the total number of curves.
\item The mean $\chi^2$ between the measured time delay $\edt$ and the true value $\Delta t_i$ weighted using the {\it estimated} uncertainties $\delta_i$,
\begin{equation}\label{eq:chi2}
\chi^2 = \frac{1}{N_{\rm submit}}\sum_i \left( \frac{\edt - \Delta t_i}{\delta_i}\right) ^2 .
\end{equation}
\item The mean ``claimed'' precision $P$ of the time-delay measurements, computed from the estimated uncertainties $\delta_i$,
\begin{equation}\label{eq:P}
P = \frac{1}{N_{\rm submit}}\sum_i \left( \frac{\delta_i}{|\Delta t_i|}\right) .
\end{equation}
\item The mean accuracy $A$ of the time-delay measurements,
\begin{equation}\label{eq:A}
A = \frac{1}{N_{\rm submit}}\sum_i \left( \frac{\edt - \Delta t_i}{\Delta t_i}\right) .
\end{equation}
\end{enumerate}
To analyse the results in greater detail, we also introduce two modified metrics. First, a ``blind'' precision,
\begin{equation}
\tilde{P} = \frac{1}{N_{\rm{submit}}}\sum_i \tilde{P}_i = \frac{1}{N_{\rm{submit}}}\sum_i \left( \frac{\delta_i}{\edt}\right),
\label{eq:guessP}
\end{equation}
where we replace in Eq.~\ref{eq:P} the true value of $\Delta t_i$ by its estimation $\widetilde{\Delta t_i}$. This metric can be computed without knowing of the true time delays; its summation terms are useful, for instance to sort light curve pairs of a real monitoring survey.
Second, we introduce a variant of the accuracy
\begin{equation}
A_{\rm{abs}} = \frac{1}{N_{\rm{submit}}}\sum_i \left( \frac{\edt - \Delta t_{\rm{i}}}{|\Delta t_{\rm{i}}|}\right),
\label{eq:Amod}
\end{equation}
where we replace the $\Delta t_{\rm{i}}$ in the denominator of Eq.~\ref{eq:A} by its absolute value. While $A$ is sensitive to a bias on the amplitude of $\widetilde{\Delta t_i}$ (i.e., over- or underestimation of delays), $A_{\rm abs}$ responds to signed biases.

Statistical uncertainties on these metrics are computed following
\begin{equation}\
\mathrm{X_{err}} = \sigma_{\mathrm{X}}/\sqrt{N_{\rm{submit}}},
\label{eq:xerr}
\end{equation}
where $\sigma_{\mathrm{X}}$ is the sample standard deviation of the summation terms of $ \mathrm{X} = \chi^2$, $P$, and $A$.

\section{\stageone : discovering time delays}
\label{sec:stage1}

To apply any of the \pycs time-delay measurement algorithms \citep{pycs} to a pair of light curves, a prior estimate of the time delay is required. Depending on the considered light curves, identifying this delay from the data might be difficult or even impossible. In the following, we describe two approaches to discover rough time-delay estimates (\stageone). Both methods rely solely on the light curves without considering the configuration of the lens system. The first method is based on a visual inspection of the light curves and is the method we used to blindly prepare submissions for the TDC1 \citep{Liao2014}. We developed the second method after the TDC1 results were released. We use the data from the challenge to evaluate the relative merits and drawbacks of each method.

\subsection{{\tt D3CS}: D3 visual curve shifting}
\label{sec:d3cs}

This method is based on visual inspection of the light curves, in the spirit of citizen science projects \cite[e.g., see the review by][]{Marshall2014}. To ease this process, we developed a dedicated browser-based visualization interface, using the {\tt D3.js} JavaScript library\footnote{Data-Driven Documents, \url{http://www.d3js.org/}} by \citet{d3}. We have made this interface public\footnote{\url{http://www.astro.uni-bonn.de/~mtewes/d3cs/tdc1/} (See ``Read me first'' for help)}.

The main motivations behind this time-costly yet simple approach were (1) to obtain rough initial estimates for the time delays and their associated uncertainties, and (2) to estimate how confident one can be that the time-delay estimations are not catastrophic outliers. Clearly, visual curve-shifting allows for more freedom than any automated procedure. It also permits dealing in a more flexible way with unusual behaviour of light curves, such as highly variable signal-to-noise from one season to the next, extreme microlensing, or even when the time delays are comparable in length to the visibility period of the object.   

\begin{figure}
\centering
\includegraphics[width=0.49\textwidth]{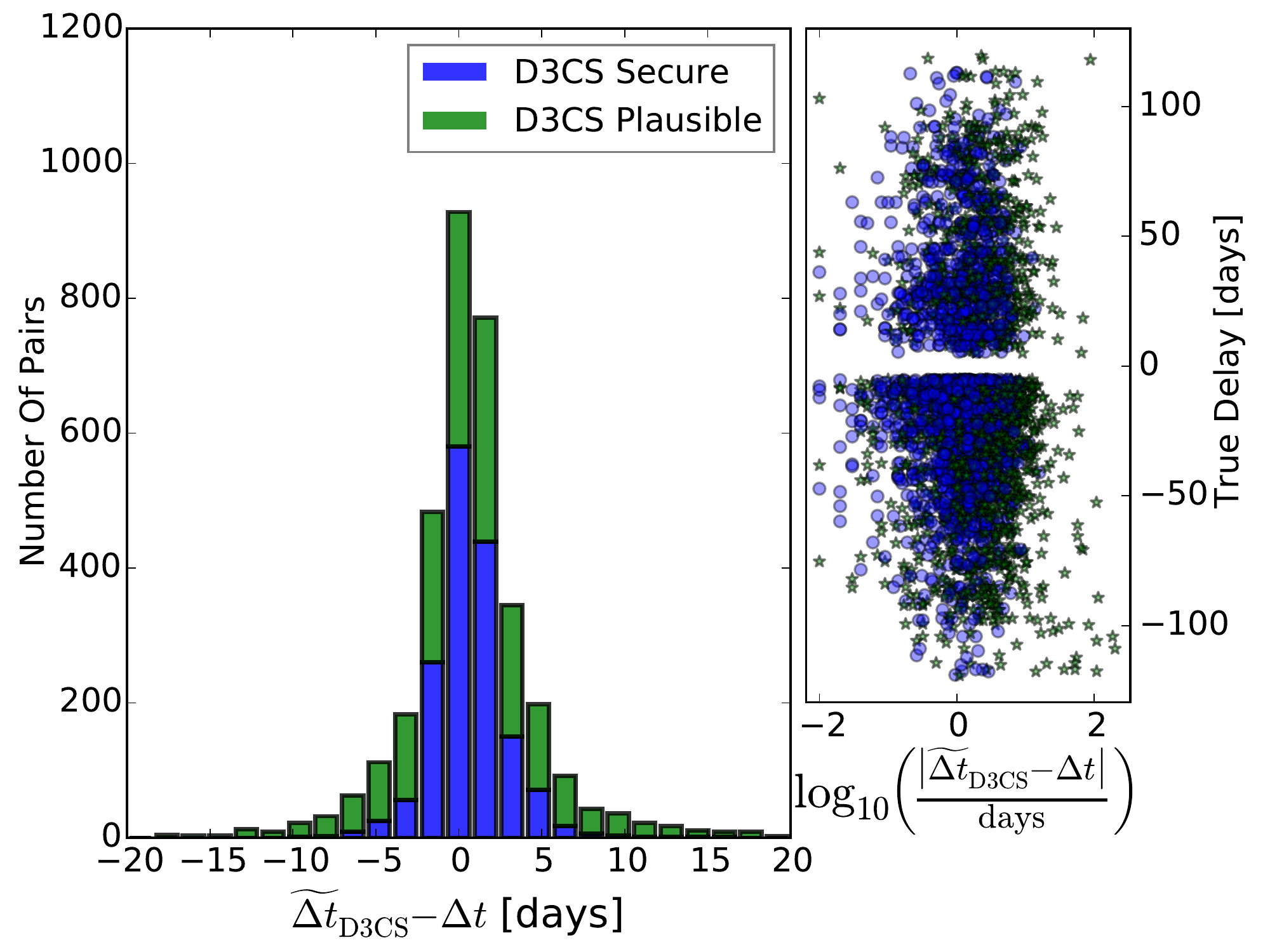}
\caption{Left panel: stacked histogram of the errors made by the visual time-delay estimation, for the {\tt secure} and {\tt plausible} {\tt D3CS} samples. All the {\tt secure} estimations are within the displayed range of errors of $\pm20$ days. Only $2.6\%$ of the time delays in the {\tt plausible} sample have an absolute error larger than 20 days. Right panel: absolute time-delay error made by D3CS as a function of the true delay for both samples.}
\label{fig:doupla_td}
\end{figure}

Our interface allows users to interactively shift the light curves in time, magnitude, and flux, and to zoom in on sections of the data. It permits the visual estimation of the time delay and of an associated uncertainty. Importantly, the interface also asks to pick one of four choices of confidence category for the proposed solution:
\begin{enumerate}
\item {\tt secure}: if a catastrophic error can be excluded with a very high confidence\footnote{This same category was named {\tt doubtless} in \citet{Liao2014}};
\item {\tt plausible}: if the solution yields a good fit and no other solutions are seen;
\item {\tt multimodal}: if the proposed solution is only one among two or more possible solutions that are equally satisfactory;
\item {\tt uninformative}: if the data does not allow the estimate of any time delay.
\end{enumerate}
Unlike massive crowdsourcing programs \cite[e.g. {\tt Galaxy Zoo;}][]{Lintott2008}, only four scientists participated in the visual inspection of TDC1 and each pair of curves was measured by at least two independent scientists. The behaviour of different users in terms of time spent per time-delay measurement span a broad range. Fast users spend on average 25 seconds per object, while slower users spend more than 60 seconds per object. This includes the time taken to measure a delay, to estimate a 1$\sigma$ uncertainty, and to allocate one of the four confidence levels described above.

To obtain a single \stageone estimation for each light curve pair, we reduce the results obtained by all four scientists in a very conservative way. We systematically downgrade to {\tt multimodal} the confidence category of pairs with conflicting time-delay estimates.

We define samples combining the four confidence categories as follows: \sec stands for {\tt secure} only, \secpla stands for {\tt secure + plausible}, and {\tt secplamul} for {\tt secure + plausible + multimodal}. The combination of all estimations is labelled {\tt all}.

Figure \ref{fig:doupla_td} shows the distribution of the error on the time-delay estimation versus the true time delay for the {\tt secure} and {\tt plausible} {\tt D3CS} subsamples. Table \ref{tab:d3cs} summarizes the results of the {\tt D3CS} classification and displays the fraction of catastrophic outliers $\epsilon$, i.e. time-delay estimations more than 20 days away from the truth. Notably, the {\tt secure} sample contains 1623 time-delay estimates free of any catastrophic outliers.

\begin{table}
\caption{{\tt D3CS} classification of the TDC1 light curve pairs.}
  \centering
 \begin{tabular}{l r r}
  \hline\hline
  {\bf Rung 0} & Estimates  & $\epsilon$ \\
  \hline
  Secure & 548 (53.5\%) & 0\% \\
  Plausible & 291 (28.4\%) & 2.1\% \\ 
  Multimodal & 60 (5.9\%) & 30.0\% \\ 
  Uninformative & 125 (12.2\%) & -- \\
 \end{tabular}
 
  \begin{tabular}{l r r}
  \hline\hline
  {\bf Rung 1} & Estimates  & $\epsilon$ \\
  \hline
  Secure & 288 (28.1\%) & 0\% \\
  Plausible & 383 (37.4\%) & 1.3\% \\ 
  Multimodal & 127 (12.4\%) & 17.3\% \\ 
  Uninformative & 226 (22.1\%) & -- \\
 \end{tabular}

  \begin{tabular}{l r r}
  \hline\hline
  {\bf Rung 2} & Estimates  & $\epsilon$ \\
  \hline
  Secure & 223 (21.8\%) & 0\% \\
  Plausible & 406 (39.6\%) & 1.2\% \\ 
  Multimodal & 168 (16.4\%) & 27.4\% \\ 
  Uninformative & 227 (22.2\%) & -- \\
 \end{tabular}   
 
 \begin{tabular}{l r r}
  \hline\hline
  {\bf Rung 3} & Estimates  & $\epsilon$ \\
  \hline
  Secure & 329 (32.1\%) & 0\% \\
  Plausible & 324 (31.7\%) & 4.9\% \\ 
  Multimodal & 161 (15.7\%) & 18.6\% \\ 
  Uninformative & 210 (20.5\%) & -- \\
 \end{tabular}
 
  \begin{tabular}{l r r}
  \hline\hline
  {\bf Rung 4} & Estimates  & $\epsilon$ \\
  \hline
  Secure & 235 (23.0\%) & 0\% \\
  Plausible & 430 (42.0\%) & 3.5\% \\ 
  Multimodal & 108 (10.5\%) & 26.9\% \\ 
  Uninformative & 251 (24.5\%) & -- \\
 \end{tabular}

  \begin{tabular}{l r r}
  \hline\hline
  {\bf All Rungs} & Estimates  & $\epsilon$ \\
  \hline
  Secure & 1623 (31.7\%) & 0\% \\
  Plausible & 1834 (35.8\%) & 2.6\% \\ 
  Multimodal & 624 (12.2\%) & 23.2\% \\ 
  Uninformative & 1039 (20.3\%) & -- \\
  \hline
 \end{tabular}  
 \tablefoot{{The \tt D3CS} visual estimates for the time delays are shown for the 4 confidence categories defined in Sect.~\ref{sec:d3cs}. The fraction of \emph{catastrophic outliers} is given for each rung by $\epsilon$, i.e., the time-delay estimations that are more than 20 days away from the truth.}
 \label{tab:d3cs}
\end{table}

Through this simple experiment, we have demonstrated that such an approach is easily manageable for about 5000 light curves. In total the four scientists involved in the visual estimation of the time delays spent 150 hours measuring the 5120 delays. We note that 30$\%$ of the time delays were measured by three or more users.

\subsection{Attempt to design an automated \stageone procedure}

Visual inspection of the light curves is a time-consuming process that cannot be repeated many times. Designing an automated method whose efficiency approaches that of {\tt D3CS} is therefore complementary and would help to minimize the time spent on the visual inspection. We developed such a method after the end of TDC1. The concept of the method is to estimate a time delay by fitting a spline on one of the two light curves, and computing the residual signal of the second light curve relative to the spline after applying time and magnitude shifts. The details of the method are described in Appendix~A; the present section evaluates its performance and compares this estimation to the visual time-delay values.

\begin{figure}[t!]
\centering
 \includegraphics[width=0.49\textwidth]{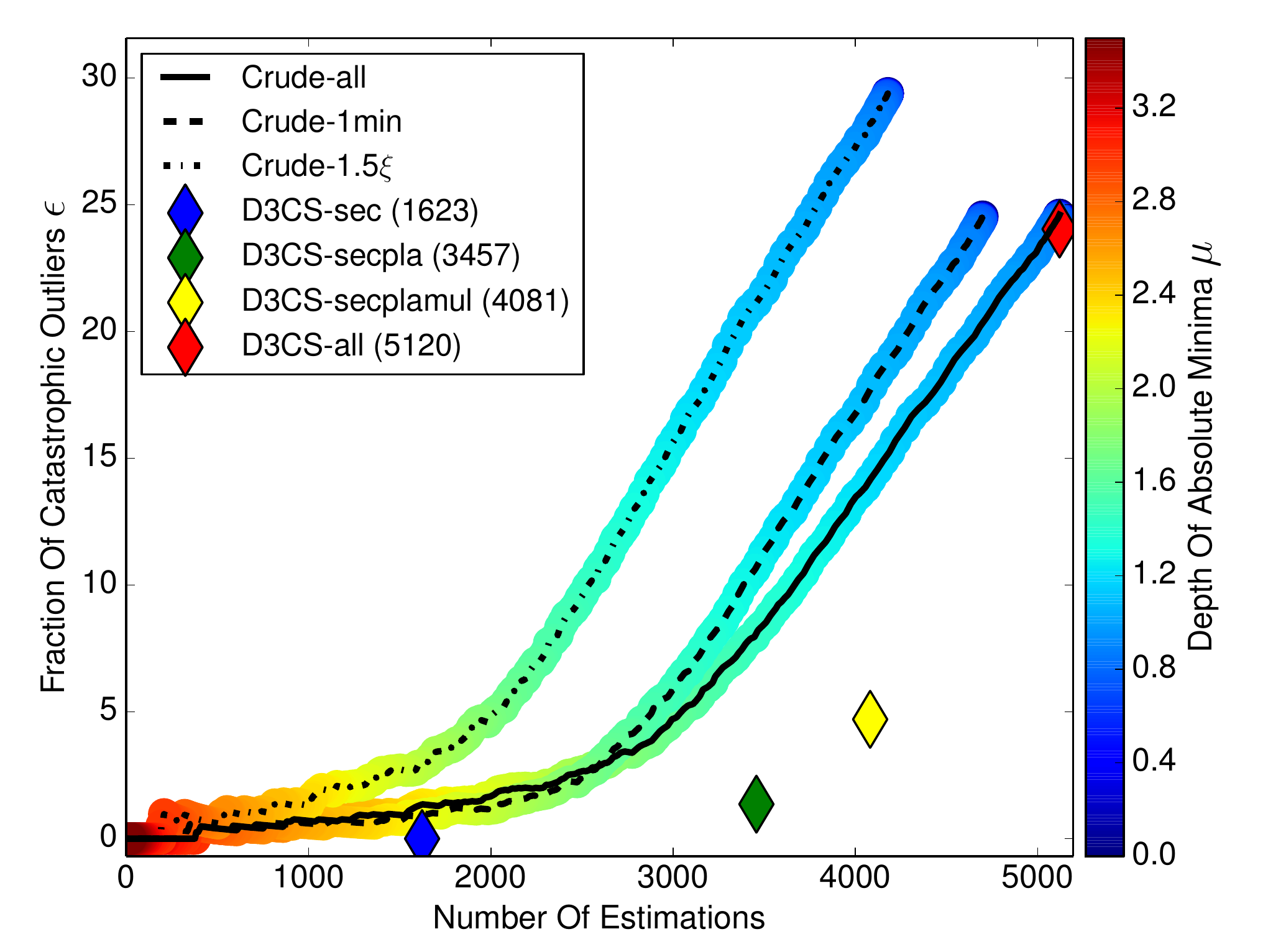}
 \caption{Cumulative evolution of the fraction of catastrophic outliers $\epsilon$ (in percentage points) as a function of the number of time-delay estimations. To produce the plot, the curves are first sorted according to the depth of their absolute minimum $\mu$, indicated in the colour bar. Each black line (solid, dashed and dotted) represents a different subsample (see text for details). The coloured diamonds show the value of $\epsilon$ for the {\tt D3CS} combined samples; the corresponding number of estimations are indicated in parenthesis.}
 \label{fig:crit_vs_ncurves}
\end{figure}

We characterize the efficiency of the automated \stageone procedure by comparing its fraction of catastrophic outliers $\epsilon$ with that of {\tt D3CS}. We define catastrophic outliers as time-delay estimations deviating from the truth by more than 20 days, i.e. with $|\edt -\Delta t_{\rm{i}}| > 20 \  \textrm{days}$. The time-delay and confidence estimation evaluated by the automated procedure are reduced to two numbers: the depth of the absolute minimum $\mu$ and the interseason variations of the microlensing $\xi$.
Figure~\ref{fig:crit_vs_ncurves} shows the evolution of the fraction of catastrophic outliers $\epsilon$ as a function of the cumulative number of time-delay estimations, sorted by increasing $\mu$. The larger $\mu$ is, the more confident the automated procedure in the time-delay estimation is. We study the impact of the automated procedure parameters $\mu$ and $\xi$ by introducing three subsamples of automated estimations:
\begin{itemize}
\item the \texttt{Crude-all} subsample contains all the estimations;
\item the \texttt{Crude-1min} subsample contains only the estimations for which the procedure finds a unique local minimum with a depth $\mu < $1;
\item the \texttt{Crude-1.5$\xi$} subsample contains only the estimations with a magnitude shift deviation $\xi < 1.5$ at the location of the absolute minimum $\mu$.
\end{itemize}
Figure~\ref{fig:crit_vs_ncurves} shows the fraction of outliers $\epsilon$ in the three subsamples and compares them to the value obtained visually with {\tt D3CS}, which are shown as four diamond-shaped symbols corresponding to the combination of the four confidence categories of {\tt D3CS} described in Sect.~\ref{sec:d3cs}. We note that the uninformative {\tt D3CS} estimations are systematically considered as catastrophic outliers here. 

The selection criteria applied to the {\tt Crude-1min} and {\tt Crude-1.5$\xi$} subsamples are not able to decrease the fraction of outliers. This highlights how difficult it is to find efficient selection criteria for the automated procedure parameters, although no exhaustive exploration of the parameters space has been conducted. As expected, all the {\tt D3CS} subsamples contain fewer outliers than the corresponding automated procedure subsamples. However, the efficiency of the latter are of the same order as the other automated methods presented in the TDC1 paper, which have $\epsilon=2-3$\% when half of the TDC1 data, i.e. 2500 light curve pairs, are measured.

In conclusion, although the automated procedure presented here is less complete and reliable than {\tt D3CS}, it yields candidates that can be evaluated by eye in a second phase. Such a combined approach would benefit both from the speed of the automated method and from the flexibility of the human eye estimate when dealing with a broad variety of properties in the light curves. We note, however, that in the rest of the paper, we only use the results obtained via {\tt D3CS} as our \stageone estimates.

\section{\stagetwo: measuring time delays}
\label{sec:stage2}

Using the \stageone results as initial estimates, we proceed in this section by running our \pycs time-delay measurement techniques on the simulated TDC1 light curves. In \citet{pycs}, three different algorithms were proposed: the simultaneous fit of the light curves using free-knot splines, the regression difference technique, and an approach based on a dispersion measurement of which the free-knot splines and the regression difference technique yielded the most accurate and precise results when applied to simulated COSMOGRAIL data \citep[in][]{Courbin2011, Tewes:2013iz, Eulaers2013, RathnaKumar:2013eu}. To analyse the TDC1 simulations, we have therefore focused on adapting only these two most promising methods for an automated use.

We note again that our \stagetwo methods cannot be asked to judge the plausibility of a proposed delay. This step belongs to the \stageone method, i.e. to the visual inspection with D3CS to prevent or at least reduce catastrophic outliers. In practice, despite a correct \stageone estimate, any automated \stagetwo method may fail to converge, or it may yield a time-delay measurement that is incompatible with the initial approximation. To prevent these failures from contaminating our measurements we systematically discard any \stagetwo result that does not lie within 1.5 {\tt D3CS} uncertainty estimate of the {\tt D3CS} point estimate. This threshold acknowledges that the uncertainty estimates obtained from {\tt D3CS} are typically overestimated by a factor of 2 to 3, which has been confirmed by \citet{Liao2014}. We note that this rejection affects less than 1\%\ of the light curve pairs and has no significant influence on the $f$ metric.

\subsection{Free-knot spline technique}
\label{sec:spl}

In the free-knot spline technique ({\tt spl}), each light curve in a pair is modelled as the sum of a spline representing the intrinsic variability of the quasar, common to both images of the pair, and an independent spline representing the extrinsic variability due to microlensing. The intrinsic spline has a higher density of knots and is therefore more flexible accomodating the quasar variability, which is assumed to be faster than the microlensing variability. During the iterative fitting process, the light curves are shifted in time so as to optimize the $\chi^2$ between the data and the model light curve. To analyse a TDC1 light curve pair, we repeat this fit 20 times, starting form different initial conditions covering the \stageone uncertainty. This tests the robustness of the optimization procedure. The best model fit is then used to generate 40 simulated noisy light curves with a range of true time delays around the best-fit solution and using the temporal sampling of the original light curves. By blindly rerunning the spline fit on these simulated data, and comparing the resulting delays with the true input time delays, the delay measurement uncertainty is estimated.

We simplified and automated the {\tt spl} algorithm for TDC1 with respect to the description of the free-knot spline method given in \citep{pycs} and its applications to real COSMOGRAIL data. The main adaptations are the following:
\begin{enumerate}
\item The temporal density of spline knots controlling the flexibility of the intrinsic spline was computed from the signal-to-noise ratios measured on the two light curves, using an empirical calibration. The signal-to-noise ratios were obtained from a structure function, by comparing the typical amplitude of the light curve variability observed on a timescale of 50 to 75 days with the scatter between temporally adjacent observing epochs. For the microlensing spline, the knot density was fixed to be the same for all TDC1 pairs.
\item When generating our mock light curves, we did not inject any fast microlensing signal to mimic correlated noise. Only plain white noise was added to the generative model. 
\item We did not analyse the time-delay measurement errors on the simulated curves as a function of true time delay. Instead, only the RMS error of these time-delay measurements was used as our total uncertainty estimate.
\item Finally, we did not manually fine-tune any parameters or correct for problematic model fits. 
\end{enumerate}
As a result, the entire {\tt spl} analysis took about 5 CPU-minutes per TDC1 pair.

\begin{figure*}[t!]
\centering
\includegraphics[width=\textwidth]{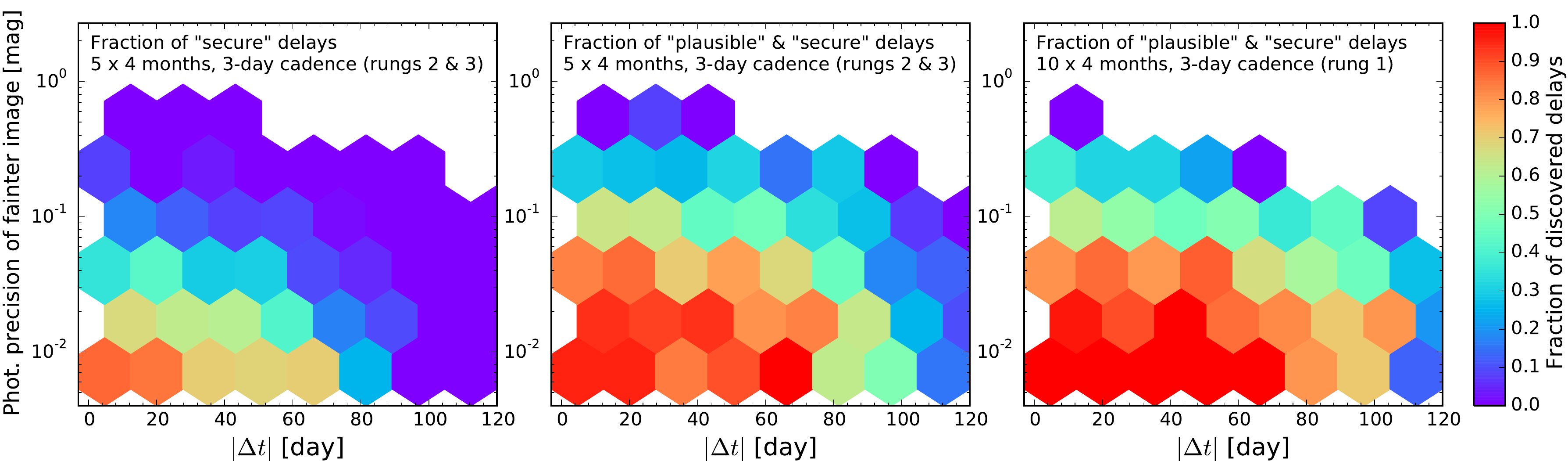} 
\caption{
Quantitative analysis of the discoverability of time delays through the extensive visual search with {\tt D3CS} (\stageone) in the case of four-month observing seasons and a cadence of 3 days. The coloured tiles show the fraction of discovered delays as a function of the photometric precision of the fainter quasar image and the true time delay of the system. The left panel shows results from the very conservative \sec sample, and the central panel shows the less pure \secpla sample that includes delay candidates considered as {\it plausible} (see text). The right panel, also for \secpla, doubles the number of observing seasons. In each panel, only tiles covering more than three light curve pairs are shown.
}
\label{fig:f}
\end{figure*}

\begin{figure*}[tbp]
\centering
 \includegraphics[width=0.80\textwidth]{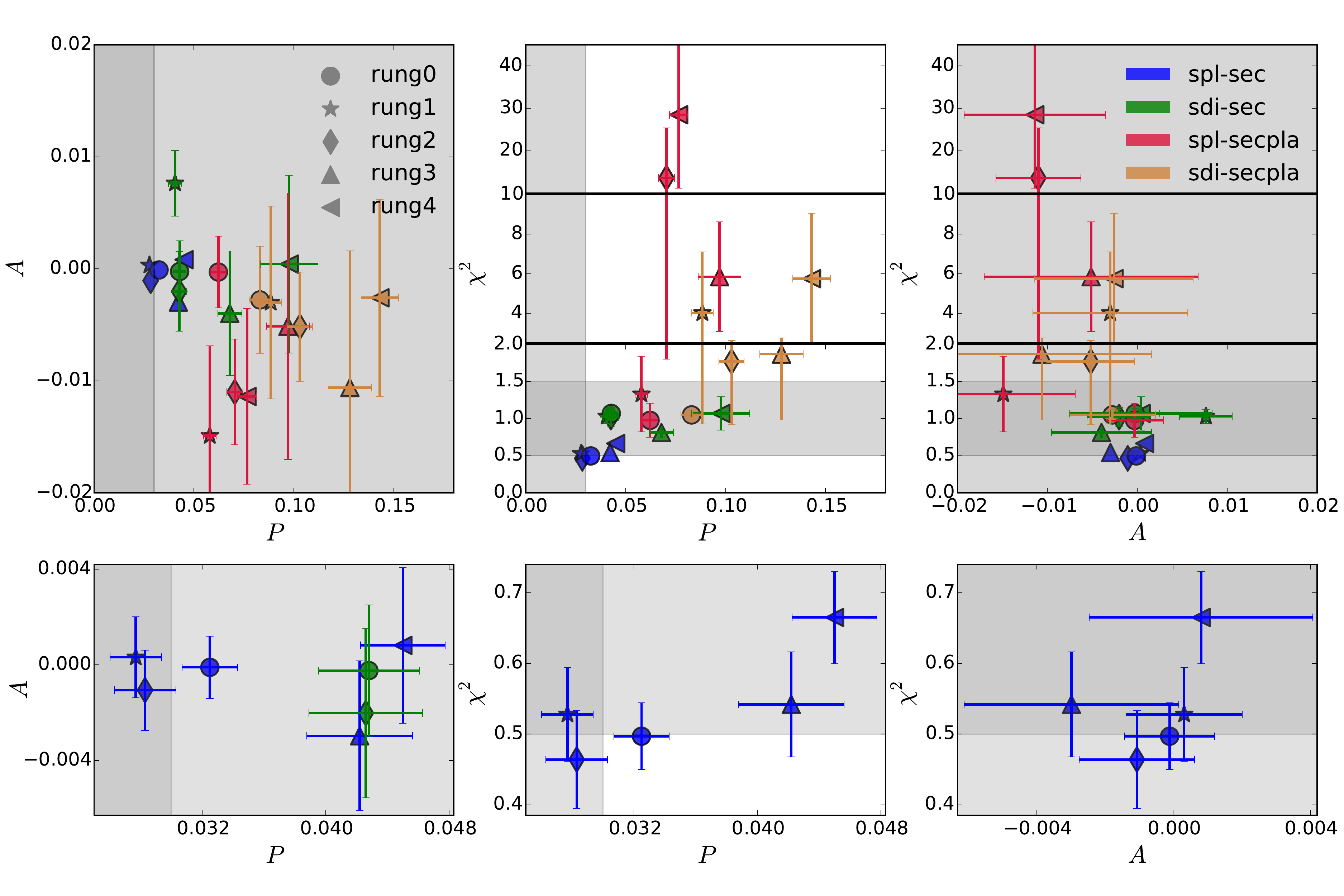} 
 \caption{Summary of metrics obtained with the \stagetwo algorithms \spl and \sdi, without any a posteriori clipping of the outliers. The bottom row presents enlargements taken from the panels on the upper row. The shaded regions represent the somewhat arbitrary target areas that were defined in the TDC1 paper.}
\label{fig:metricsresults}
\end{figure*}

\subsection{Regression difference with splines}\label{sec:sdi}

Our second \stagetwo method, {\tt sdi} (for spline difference), is based on the regression difference technique of \citet{pycs}. To speed up the analysis, we replace the Gaussian process regressions by spline fits. In summary, the method independently fits a different spline to each of the two light curves, and then minimizes the variability of the difference between these two splines by shifting them in time with respect to each other. The advantage of this approach is that it does not require an explicit microlensing model. To estimate the uncertainty, the {\tt sdi} method is run on the same simulated light curves provided by the {\tt spl} technique. The {\tt sdi} method has an even lower computational cost than {\tt spl}.

\section{Results on the Time Delay Challenge 1 (TDC1) data}\label{sec:results}

In this section, we separately analyse results from the \stageone and \stagetwo measurement processes as obtained on the simulated light curves of TDC1. General results from \citet{Liao2014} regarding submissions prepared with our methods include the following:
\begin{enumerate}
\item The \stagetwo methods \spl and \sdi show no significant deviations of the accuracy $A$ from zero, and can thus be considered as inherently unbiased, given the statistical limits due to the finite challenge size.
\item The claimed precisions $P$ of \spl and \sdi are very good, with $\chi^2$ values of the order of $\chi^2_{\rm \spl}\simeq0.5$ and $\chi^2_{\rm \sdi}\simeq1.0$.
\item Based on results from the $\spl$ method simple power-law models for the dependence of $A$, $P$, and $f$ on monitoring cadence, season length, and campaign length were adjusted. These relations attempt to capture general trends regarding the behaviour of all methods used in the context of TDC1, including our \spl technique.
\end{enumerate}
In the present paper, we focus on aspects that are complementary to the discussion of \citet{Liao2014}.

\begin{figure*}[tbp]
\centering
\includegraphics[width=0.9\textwidth]{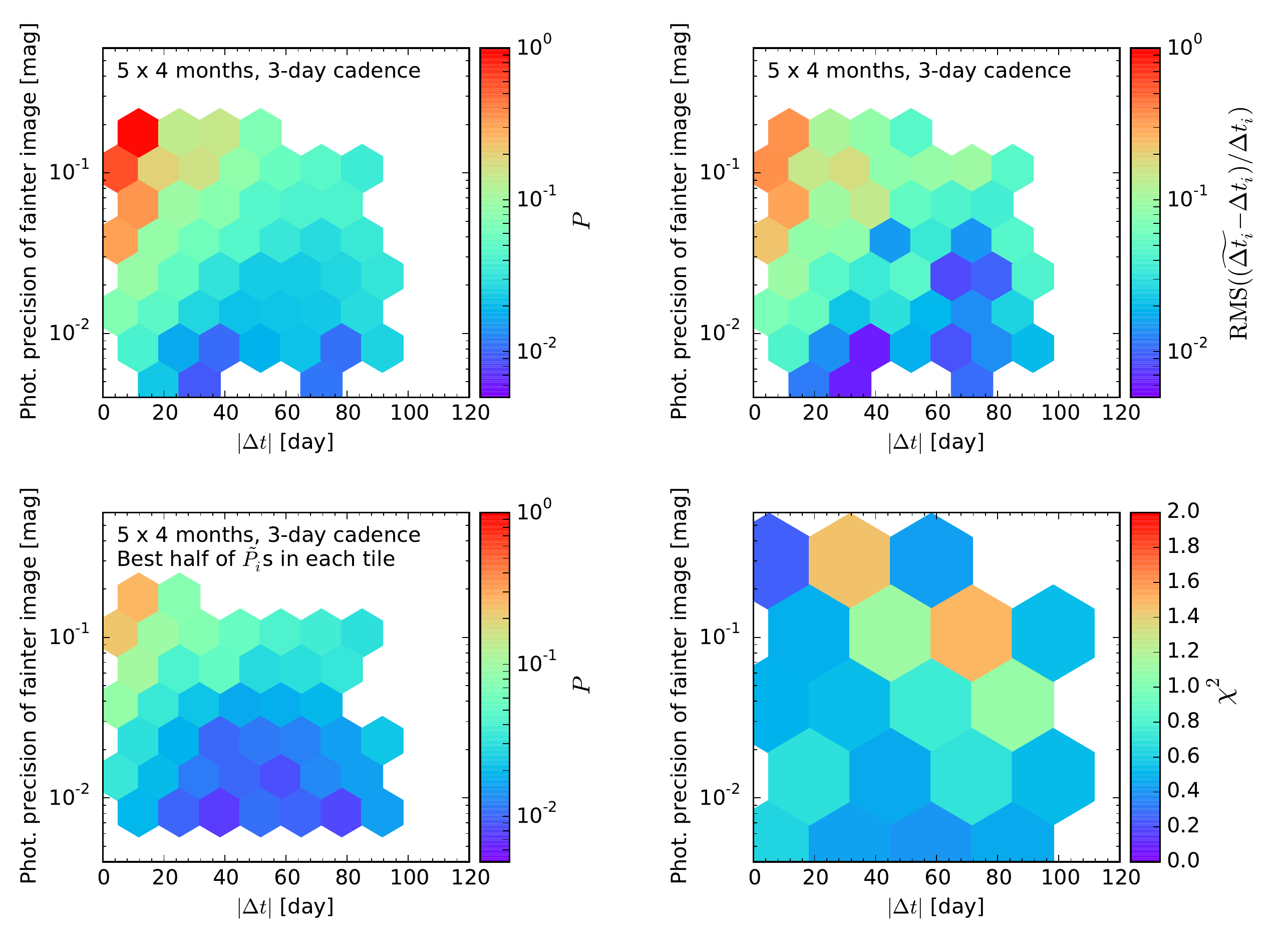} 
\caption{
Quantitative analysis of the precision achieved for the \stagetwo time-delay measurement as a function of the photometric precision of the fainter quasar image and as a function of the true time delay. All panels show results from the bias-free \spl technique for the {\tt secure + plausible} selection of rungs 2 and 3 after rejection of the catastrophic outliers (see text). The top left panel shows the metric $P$ as computed using the uncertainty estimates $\delta_i$ without using $\widetilde{\Delta t_i}$. The top right panel shows the RMS of the relative point estimation residuals without considering $\delta_i$. The bottom left panel shows the average $P$ obtained in each tile after selecting only the best half of systems according to the blind precision in $\tilde{P}$. The bottom right panel shows a map of the $\chi^2$ metric. In all panels, only tiles describing more than three light curve pairs are shown.}
\label{fig:rms_p}
\end{figure*}

\subsection{Efficiency of time-delay discovery (\stageone)}
\label{subsec:f}

We start by analysing the fraction of light curve pairs for which a time delay can be {\it discovered} with visual inspection, as a function of time delay and image brightness. This analysis relates only to the first stage of the time-delay measurement process.

Aside from the time delay and the quasar image brightness, the question of discoverability depends on observational conditions (e.g. monitoring cadence and duration) and on astrophysical characteristics (e.g. amount of quasar variability and microlensing perturbations). In the following, we select a given observing strategy and average over the astrophysical parameters of the TDC1 simulations, which follow clearly motivated distributions \citep{Dobler2013}. A large sample of light curve pairs with almost identical observing conditions can be obtained by merging rungs 2 and 3. These rungs share the fiducial three-day monitoring cadence for five seasons of four months each. The differing cadence \emph{dispersion} of $0.0$ days for rung 2 and $1.0$ days for rung 3 \citep[Table 1 of][]{Liao2014} do not have a significant impact on the discoverability of time delays. 

In practice, time delays can be measured accurately in pairs of light curves if the quality of both light curves is sufficient. In the following we consider as a relevant observable the photometric precision achieved in the fainter image of a pair. This is computed for each pair of  light curves, as the median of the photometric error bars across the epochs of the TDC1 simulations. This is made legitimate by their overall effectiveness in representing the amplitude of the simulated noise, except for very few ``evil'' epochs of some systems \citep[see Section 2.5 of][]{Liao2014}. However, when analysing real light curves, using the photometric scatter between the points might be a better choice than using potentially mis-estimated photometric error bars.

Figure \ref{fig:f} presents the distribution of the fraction of light curve pairs for which time delays could be identified via a meticulous {\tt D3CS} visual inspection for two different monitoring strategies. In the left panel, only time delays categorized as \emph{secure} through the visual inspection are considered as discovered. This is very conservative because in a real survey, simple lens models will help to identify the correct time delay for almost all of the {\it plausible} systems as well. For this reason we also consider the combined \secpla sample, shown in the central panel. 

Some of the cases categorized as {\it multimodal} could certainly also be resolved using simple lens model considerations, but in practice the vast majority of these light curve pairs contain too few clear common features to estimate a reliable time delay, even if an approximate value would be known from the modelling. We therefore consider the discoverability of the \secpla selection shown in the central panel of Fig. \ref{fig:f} as roughly representative of the fraction of potentially helpful delays that can be reliably measured from a real monitoring campaign or survey. It can be seen as an approximate lower limit for the fraction of time delays that can be observationally constrained in the absence of prior from a lens model, provided the properties of the microlensing signal are similar to those of the simulated light curves used here. Finally, the right panel shows the evolution of this discoverability if the same monitoring strategy is carried out for five more seasons, i.e., for a total of ten years.

We observe that after five years of monitoring, light curve pairs with a photometric precision in the fainter image better than $\sigma=0.1$ mag and a true time delay shorter than $\Delta t = 80$ days (2/3 of the season length) are very likely to yield a time-delay measurement. Pursuing the monitoring for five more years significantly increases the average chances that longer delays up to $\sim 90\%$ of the season length become measurable.

\begin{figure*}[htb]
\centering
 \includegraphics[width=0.90\textwidth, angle=0, origin=c]{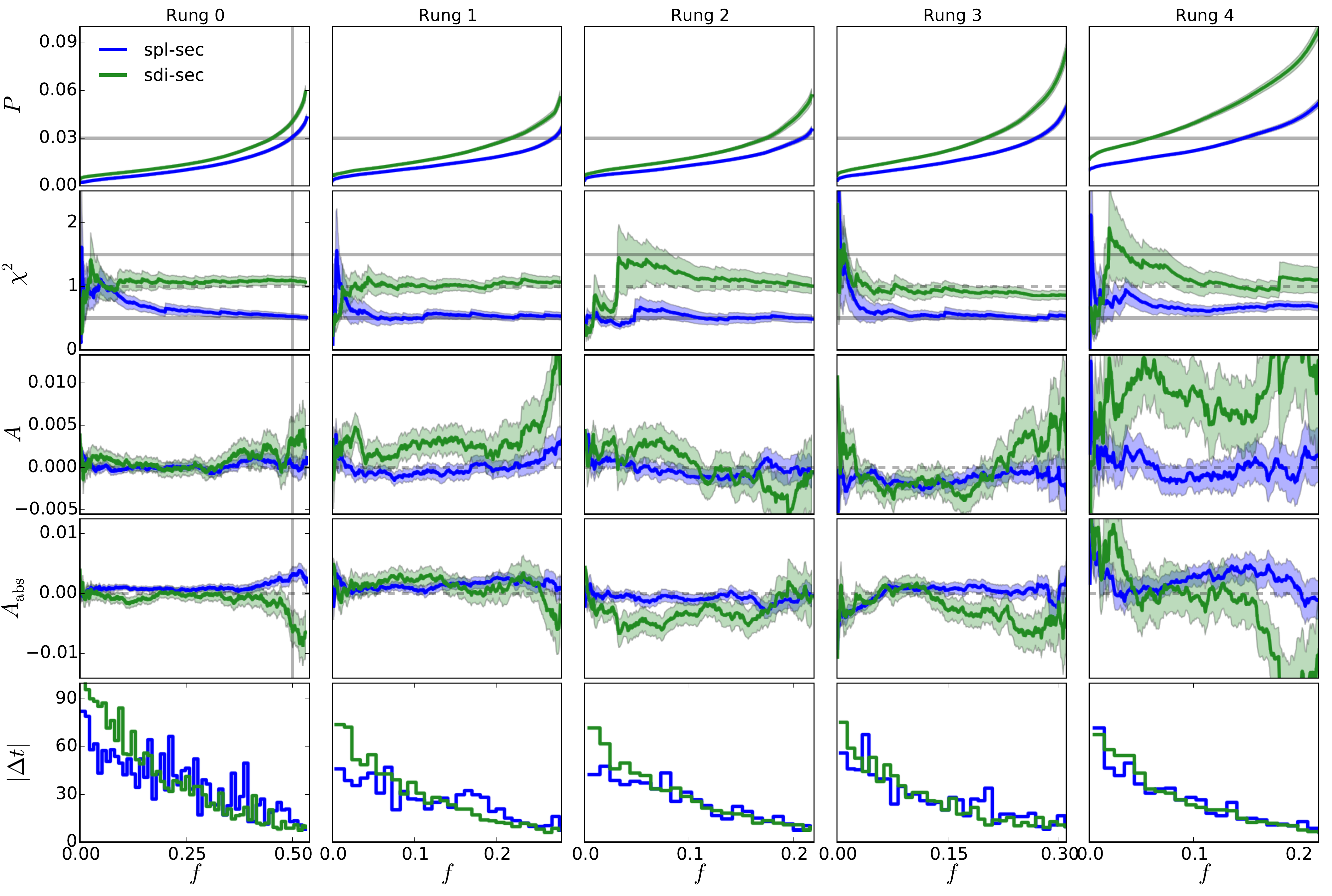}  
 \caption{Evolution of the TDC1 metrics per rung as a function of the fraction of estimations $f$ for the \splsec and \sdisec samples. The plots are sorted by increasing values of the blind precision $\tilde{P}$ (see text). Shaded regions represent the error on the metrics. Solid grey lines show the target values for the metrics as defined in the TDC1 paper. Dashed grey lines show the best possible value for each metric. The bottom row presents the non-cumulative evolution of the median of the {\it true} time delays $|\Delta t_{\rm{i}}|$ in bins of ten estimations.}
 \label{fig:XX-vanilla-dou-full}
\end{figure*}

\subsection{Precision of time-delay measurement (\stagetwo)}

We now turn to the analysis of the time-delay measurements (\stagetwo) for all systems where the time delay is successfully discovered (\stageone).

Figure \ref{fig:metricsresults} summarizes the results of the \spl and \sdi methods in terms of the metrics $A$ (accuracy), $P$ (claimed precision), and $\chi^2$, as defined in Sect.\ref{sec:tdc}. The figure follows the same conventions as Table 4 of \citet{Liao2014}, but also includes measurements obtained on the \secpla samples of each rung. As expected, the results for these \secpla samples are more scattered than for the \sec samples. The reason for these significant offsets in $A$ and $\chi^2$ with respect to the \sec results is the stronger impact of outliers on the non-robust metrics.

To characterize the achievable precision of the \stagetwo measurements without being influenced by \emph{catastrophic} outliers but still benefiting from a large number of light curve pairs, we now focus on the \secpla sample from which we remove systems with estimated delays that are off by more than 20 days. This rejects about 1\% of the \secpla sample. We also note that catastrophic outliers are errors of the \stageone estimate, not \stagetwo.

Figure \ref{fig:rms_p} presents metrics related to the \stagetwo time-delay measurement precision as a function of the photometric quality of the fainter quasar light curve and the time delay. In each tile the top left panel shows the average claimed precision $P$ for the \spl technique, for a five-year monitoring with four-month seasons and a cadence of three days. We find that the cadence dispersion plays no significant role in this analysis, and we therefore merge rungs 2 and 3 to obtain this larger sample.

In contrast, each tile of the top right panel shows the root mean square (RMS) of the relative error of the time-delay estimates $\widetilde{\Delta t}$. The observed structure is inevitably noisier because this RMS is computed from the actual point estimates, while the precision $P$ is based on the corresponding claimed uncertainty estimates. We observe both a qualitative and a quantitative similarity between these plots, suggesting that the time-delay uncertainty estimates, $\delta_i$ (Eq. \ref{eq:chi2}), from the \pycs techniques correctly capture trends related to the photometric quality and to the amount of temporal overlap in the light curves.

In the lower right panel of Fig. \ref{fig:rms_p}, the map of $\chi^2$ metrics for the \spl technique shows no strong evolution across the well-sampled regions of the parameter space. It does however indicate that the uncertainty estimates $\delta_i$ from \spl are too conservative by a small factor of $(\chi^2)^{-1/2}\simeq 0.5^{-1/2} \simeq 1.4$. This is particularly visible for the high quality light curves with small time delays, i.e. with large temporal overlaps. Finally, the bottom left panel shows the average $P$ metric computed using only the best half of light curve pairs in each tile, where the quality of a system is judged via the blind relative precision $\tilde{P}_i$ (see Eq. \ref{eq:guessP}). This operation, aimed at increasing the precision, divides by two the usable fraction of systems as given in Fig. \ref{fig:f}. We consider such a selection in more detail in Sect. \ref{subsec:select}.

We observe in Fig.~\ref{fig:rms_p} that the best \emph{average} relative precision in time-delay measurements seems to be achieved for time delays in the range 40-80 days for this particular monitoring strategy. This corresponds to about half of the season length, and results from the trade-off between absolute delay length and amount of overlap in the light curves.

Given the observed general aptitude of our time-delay uncertainty estimates, and thus $P$, to describe the actual point estimate errors committed by the \spl technique, and the excellent competitiveness of \spl compared to other time delay measurement techniques \citep[see, e.g., Fig. 13 of][]{Liao2014}, we see the left panels of Fig.~\ref{fig:rms_p} as roughly representative of the precision that can be achieved by a state-of-the-art time-delay measurement method.

\begin{figure}[tb]
\centering
 \includegraphics[width=0.47\textwidth]{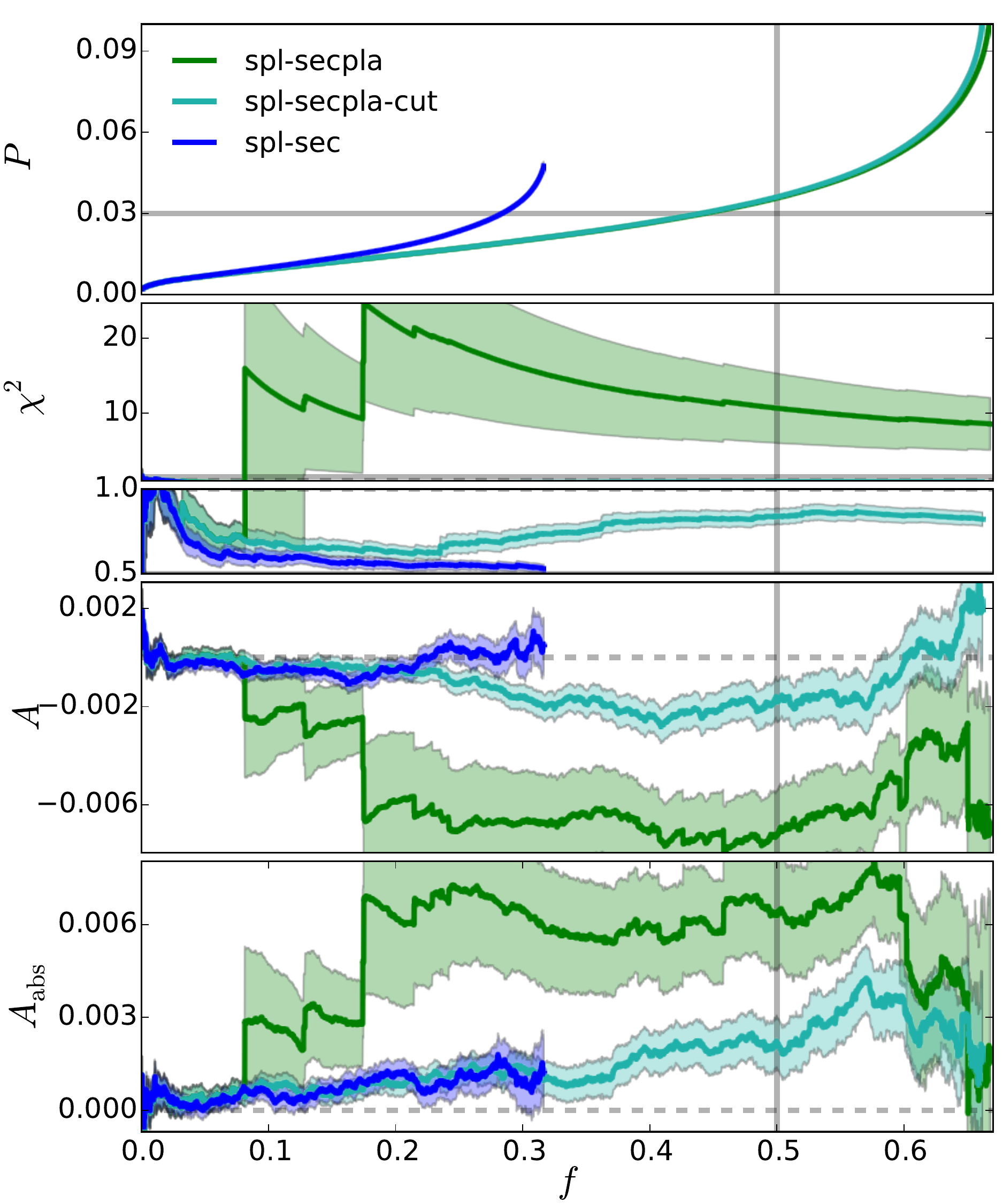} 
 \caption{Evolution of the TDC1 metrics with the fraction of estimations sorted by increasing blind precision $\tilde{P}$, for the \splsec and {\tt spl-secpla} samples merging all rungs. The {\tt spl-secpla-cut} sample has been cleaned {\it a posteriori} from the outliers in the {\tt spl-secpla} sample. In doing so, we removed 29 curves from the {\tt spl-secpla} sample. The shaded regions, the solid and dashed grey lines are the same as in Fig.\ref{fig:XX-vanilla-dou-full}.}
 \label{fig:spl-vanilla-XX-full}
\end{figure}

\subsection{Effect on the overall metrics of selecting the best systems}
\label{subsec:select}

In Fig.~\ref{fig:XX-vanilla-dou-full} we show the evolution of the overall average metrics as a function of the fraction of estimations $f$ for the \splsec and \sdisec samples. The five columns represent the five rungs and the estimations are sorted according to their blind precision $\tilde{P}$, i.e the precision estimate from the TDC1 data prior to the unblinding of the true values. The non-cumulative median value of the true delays (bottom row) is computed on consecutive bins of ten estimations. The shaded regions represent the statistical uncertainties of the metrics $\mathrm{X_{err}}$ defined in Eq.{\ref{eq:xerr}}. These uncertainties are too small in the top row to be distinguished from the curves.

In the top row, $P$ increases monotonically with $f$. This is expected since the estimations are sorted according to $\tilde{P}$ and since the {\tt D3CS sec} sample is free of any outliers. The metrics $\chi^2$, $A$, and $A_{\rm{abs}}$, respectively in the second, third and fourth rows stabilize around a mean value once a significant fraction of estimations is considered. The variations around the mean such as the jump in $\chi^2$ at $f\sim0.05$ in rung 2 are due to non-catastrophic outliers, i.e. time delays that deviate significantly from the truth but by less than 20 days. These outliers are the result of a non-optimal convergence of the \stagetwo methods for the curves with the lowest signal-to-noise.

The high-$f$ end of $A$ and $A_{\rm{abs}}$ are subject to strong variations in all rungs. These variations occur for small absolute values of the true time delay $|\Delta t|$. Similarly, the high-$f$ end of $P$ increases strongly. A small error on the time-delay estimation particularly affects $P$, $A$, and $A_{\rm{abs}}$ if the true time delay is small.

This correlation between the loss in precision and accuracy means that for the corresponding light curves,
our algorithms estimate the time delays less accurately, but do provide larger error bars. We observe that the $\chi^2$ remains constant as $f$ increases. In conclusion, sorting the measurements in $\tilde{P}$ and rejecting a small fraction of the least precise estimations allows an optimal accuracy to be maintained without affecting the $\chi^2$.

Figure \ref{fig:spl-vanilla-XX-full} shows the evolution of the TDC1 metrics with the fraction of estimations, $f$, sorted by increasing order of $\tilde{P}$, for the {\tt spl-sec} and {\tt spl-secpla} sample. The few catastrophic outliers result in striking discontinuities in the curves. Quantifying the accuracy and precision of \stagetwo methods is different from avoiding such catastrophic outliers, and to address the former question, we also display in Fig. \ref{fig:spl-vanilla-XX-full} a new subsample, {\tt spl-secpla-cut}, where the 29 time-delay estimates with an absolute time-delay error larger than 20 days are a posteriori removed. Similarly, the impact of outliers can be reduced by considering the median of the individual metrics instead of their mean. This is not surprising, but nevertheless it reflects the need either to use metrics that can cope with outliers or, as in our \stageone approach, to make sure that no outliers contaminate the time-delay samples used for subsequent cosmological application.

\section{Summary and conclusions}
\label{sec:conclusion}

In this work, we used the simulated light curves of the Time Delay Challenge 1 (TDC1) proposed by \citet{Dobler2013} to test the performance of the \pycs numerical techniques currently in use in COSMOGRAIL to measure time delays. These methods are fully data-driven, in the sense that they do not attempt to include any physical model for microlensing or for the intrinsic variations of quasars. This choice was deliberate and considers an empirical representation of the data that minimizes the risk of bias due to choosing the wrong physical model. The price to pay is that error bars on the measurements must be estimated from mocks and that we cannot use prior knowledge from external observations of quasars in a direct formal way. 
Using the same simulated light curves, we also assessed the quantity and quality of the time-delay measurements from future monitoring campaigns or surveys. We have made public our six main TDC1 submissions, obtained using the {\tt D3CS}, \spl, and \sdi methods for the high purity {\tt secure} and the less conservative {\tt plausible} samples. These data are available on the COSMOGRAIL website\footnote{\url{http://www.cosmograil.org/tdc1}}. Our results can be summarized as follows:

\begin{enumerate}

\item The visual estimation of time delays (\stageone) is extremely efficient in spotting catastrophic outliers and in providing useful time-delay estimates to be refined with subsequent numerical techniques (\stagetwo).
 
\item We attempted to build a simple automated time-delay estimation procedure that we can apply to the TDC1 data. While useful, this automated procedure does not achieve as good purity in the time-delay sample as the visual estimation. We note that we did not use this automated procedure for any of our submissions to the TDC1. 

\item We provide a general analysis of the achievability of time-delay measurements as a function of the photometric precision of the light curves. In particular we show that almost all time delays shorter than two-thirds of the season length can be measured in five years of monitoring with four-month seasons and realistic photometric quality.

\item Our \stagetwo methods \spl and \sdi can be considered unbiased given the statistical limits due to the finite challenge size. The $\chi^2$ values are close to unity. These good results emphasize the reliability of COSMOGRAIL time-delay measurements in general.

\item We quantify the average precision on the time-delay measurements as a function of photometric quality of the light curves. We find that the best average precision seems to be obtained for pairs whose time delay is approximately half of the monitoring season length.

\item We show that we can reliably evaluate the individual precision of our time-delay estimates. This may enable us, for any sample of light curves, to identify which are the most valuable objects to be followed up for cosmological purposes. We note, however, that any selection on the time delays in a quasar sample may also result in biases on the cosmological inference.

\end{enumerate}

The above is true for the specific light curves of TDC1. These curves have been generated with simple models for quasar variations and microlensing and they do not include all potential nuisances of astrophysical, atmospheric, or instrumental origin. In addition, the \pycs techniques currently used in COSMOGRAIL do not attempt to account for these effects. 

\begin{acknowledgements}
The authors would like to thank the organizers of the Time Delay Challenge, as well as Peter Schneider for useful discussions and the anonymous referee for his/her useful comments. 
This work is supported by the Swiss National Science Foundation (SNSF). Malte Tewes acknowledges support by a fellowship of the Alexander von Humboldt Foundation and the DFG grant Hi 1495/2-1. Dominique Sluse acknowledges support from a {\it {Back to Belgium}} grant from the Belgian Federal Science Policy (BELSPO). \end{acknowledgements}

\bibliographystyle{aa}
\bibliography{paper}

\clearpage

\appendix

\section{Automated \stageone procedure}\label{app:automated}

If more time had been devoted to the visual inspection, we expect that more correctly estimated {\tt plausible} time-delay estimations would have been classified as {\tt secure}. After the TDC1, we developed an automated \stageone procedure. Its goal is to speed up and possibly improve the quality of the {\tt D3CS} output by providing a range of reasonable initial time delays and associated confidence estimation.
The following section describes technically how the time-delay and confidence estimation for this automated procedure are computed.

\subsection{Time-delay estimation}

For each pair of curves, a bounded-optimal-knot spline \citep{Molinari2004, pycs} $s(t)$ is fitted to one of the two light curves. The second light curve $l(t)$ is then shifted by an amount $\delta t$ in time and $\delta m$ in magnitude. Thus, for a given observing epoch $t$, the value of the shifted light curve can be written as $l(t-\delta t)_{\rm{\delta m}}$. For every value of the time and magnitude shifts, we select all the $N$ points in the second light curve that overlap in time with the spline. For these points, we compute the residuals $\rm{res_n}$ relative to the spline, i.e. the difference in magnitude between the points and the spline. The residual ${\rm res_n}$ for point $n$ is

\begin{equation}
{\rm res}_{\rm n} = [s(t) - l(t-\delta t)_{\rm{\delta m}}]_{\rm n}.
\end{equation}

We then compute the average absolute residual $r(\delta t,\delta m)$ for every time and magnitude shift, i.e.

\begin{equation}
r(\delta t,\delta m) = \frac{1}{N}\sum_{n=1}^{N}\frac{|\rm{res_{n}}|}{\sqrt{N}}.
\end{equation}

The possible presence of microlensing, assumed to be constant over an observing season, is handled in a very simple way. For each time shift $\delta t_{\rm{i}}$, we apply independent magnitude shifts $\delta m_{\rm{j}}(\delta t_{\rm{i}})$ to each season $j$. We define the residual curve $\vec{r} = \{r_{\rm{1}},...,r_{\rm{i}},...,r_{\rm{T}}\}$ as the sum of the smallest average absolute residuals for each season $j$. The $i$ index runs from 1 to $T$ and  denotes the different time-delay values $\delta t_{\rm i}$ we want to test. This translates into

\begin{equation}
r_{\rm i} = \sum_{\rm{j}}\min_{\rm{j}}{\big{[}r(\delta t_{\rm{i}}, \delta m_{\rm{j}})\big{]}}.
\end{equation}

\begin{figure}
\centering
\includegraphics[width=0.49\textwidth]{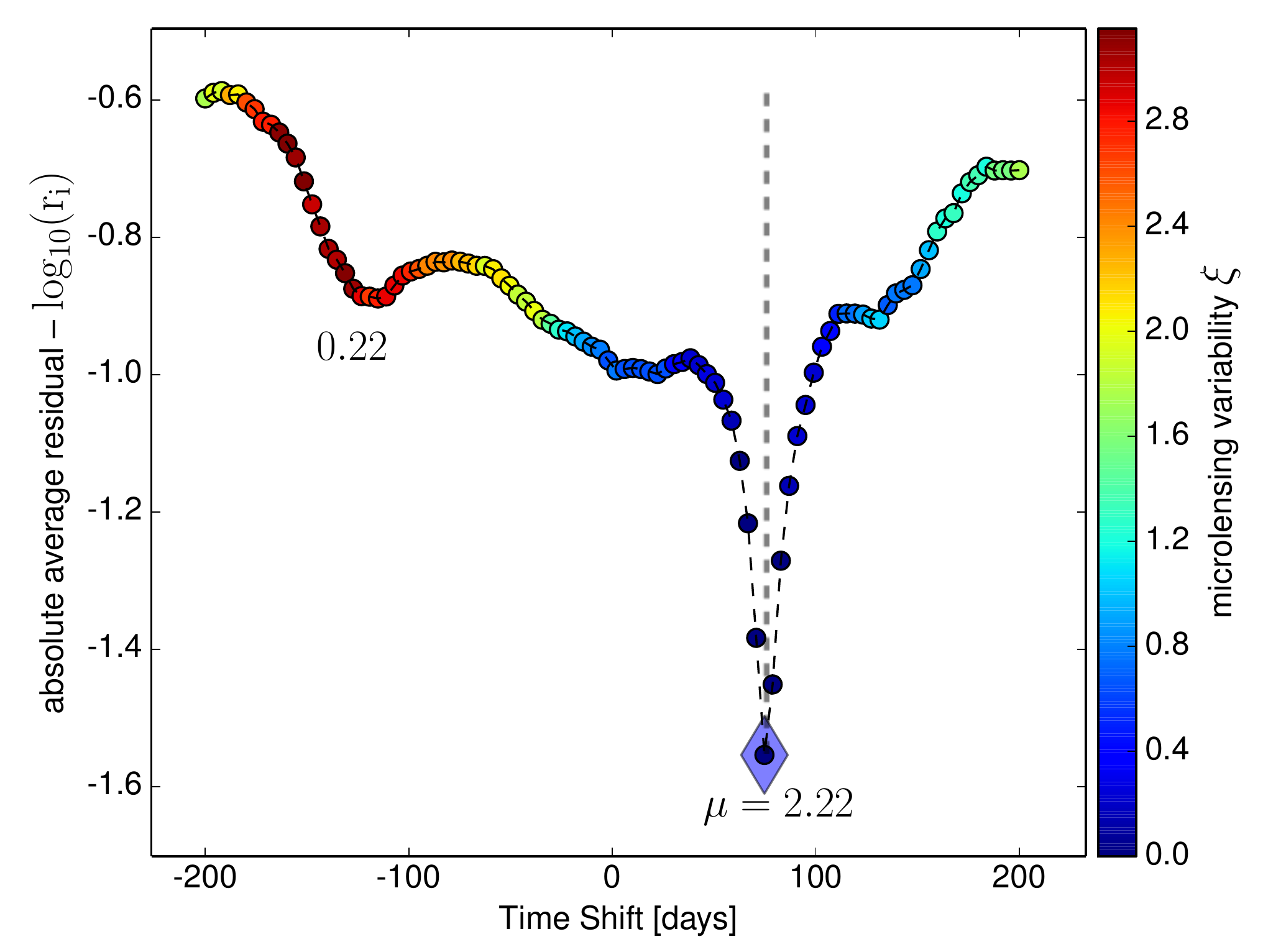}
\caption{Example of time-delay estimations and confidence level from the automated procedure. The vertical grey dashed line represents the true value of the time delay. The blue diamonds correspond to the smallest absolute average residuals $r_{\rm{i}}$ returned by the automated procedure. The depth of the two minima $\mu_{\rm{i}}$ are represented by the number below each minimum. The colour bar indicates the microlensing variability $\xi$ (see text).}
\label{fig:res_perseasons_sums}
\end{figure}

We define $r_{\rm{i}}$ as a local minimum in $\vec{r}$ if 

\begin{equation}
r_{\rm{i}} < r_{\rm{i\pm k}}, \ \ \ \rm{for} \ \ k=1...10,
\end{equation}

where we keep the absolute minimum in $\vec{r}$ as the final time-delay estimation. The $k$ index running from 1 to 10 spans a range of $\pm$ 20 days around each tested value $r_{\rm{i}}$.  Figure~\ref{fig:res_perseasons_sums} shows a typical residual curve $\vec{r}$ with the absolute minimum indicated as a coloured diamond and the true time delay indicated as a vertical dashed grey line.

\subsection{Confidence estimation}\label{sec:confidence_est}

For each pair of curves, we compute three parameters related to the shape of the residual curve $\vec{r}$ that can be used to estimate the quality of the time-delay estimations:
\begin{enumerate}
\item The number of local minima in $\vec{r}$.

\item The depth of each minimum $\mu_{\rm{i}}$ and the absolute (i.e. the deepest) minimum $\mu$,
\begin{equation}
\mu_{\rm{i}} = \frac{\bar{\vec{r}}-r_{\rm{i}}}{\sigma_{\vec{r}}} \ \ \ \ , \ \ \ \mu=\min_{\delta t_{\rm{i}}}{[\mu_{\rm{i}}]},
\end{equation}

where $\sigma_r$ is the standard deviation between the elements in $\vec{r}$. Examples of values for $\mu_{\rm{i}}$ are indicated in Fig.\ref{fig:res_perseasons_sums} below the minima for time shifts $\delta t \simeq -120$ days and $\delta t \simeq +80$ days.

\item The total magnitude shift $ \vec{\delta m} = \{\delta m_{\rm{1}},...,\delta m_{\rm{T}} \}$ and the microlensing variability $\xi(\delta t_{\rm{i}},\vec{\delta m})$, where we use the per season magnitude shifts $\delta m_{\rm{j}}(\delta t_{\rm{i}})$ minimizing the average absolute residual $r_{\rm{i}}$ at each time shift $\delta t_{\rm{i}}$, 

\begin{equation}
\delta m_{\rm{i}} = \sum_{\rm{j}} \delta m_{\rm{j}}(\delta t_{\rm{i}}) \ \ \ \ , \ \ \ \ \xi(\delta t_{\rm{i}},\vec{\delta m}) = \frac{\min{[\vec{\delta m}]} - \delta m_{\rm{i}}}{\sigma_{\vec{\delta m}}},
\end{equation}

where $\sigma_{\vec{\delta m}}$ is the standard deviation between the elements in $\vec{\delta m}$. In other words the quantity $\xi(\delta t_{\rm{i}},\vec{\delta m})$ is the difference between the sum of the magnitude shifts applied to each season at a given time shift $\delta t_{\rm{i}}$ and the minimum of this sum on all time shifts. This quantity follows the colour code in the sidebar of Fig.\ref{fig:res_perseasons_sums} and is equivalent to the season-to-season microlensing variations minimizing the residuals for a given time shift. The lower this quantity is, the smaller the impact of microlensing is.
\end{enumerate}

\end{document}